\documentclass[letterpaper, 10 pt, conference]{ieeeconf}  %

\IEEEoverridecommandlockouts                              %

\usepackage{amsmath} %
\usepackage{amssymb}  %
\usepackage[english]{babel}
\usepackage{algorithm}
\usepackage{algpseudocode}
\usepackage{pgfplots} 
\usepackage {ACSD-CR}
\usetikzlibrary{pgfplots.statistics}

\let\labelindent\relax %
\usepackage{enumitem} %

\usepackage{xspace}
\usepackage{soul}
\usepackage{mathtools}
\usepackage{makecell}
\usepackage{cite}
\usepackage{nicefrac}
\usepackage{float}
\usepackage{wrapfig}
\usepackage{color}
\usepackage{soul}

\newcommand{\ra}{\rightarrow}

\newtheorem{asm}{Assumption}
\newtheorem{prp}{Proposition}
\newtheorem{lem}{Lemma}
\newtheorem{thm}{Theorem}
\newtheorem{rem}{Remark}

\newcommand{\ie}{\unskip, i.\,e.,\xspace}
\newcommand{\eg}{\unskip, e.\,g.,\xspace}

\newcommand{\N}{\ensuremath{\mathbb{N}}}
\newcommand{\Z}{\ensuremath{\mathbb{Z}}}

\newcommand{\R}{\ensuremath{\mathbb{R}}}

\let\emptyset\varnothing

\newcommand{\bs}[1]{\ensuremath{\boldsymbol{#1}}}

\newcommand{\eps}{\ensuremath{\varepsilon}}

\DeclareMathOperator*{\argmin}{arg\,min}

\definecolor{dgreen}{rgb}{0.0, 0.5, 0.0}
\newcommand{\green}[1]{\textcolor{dgreen}{#1}}

\makeatletter
\newcommand{\subalign}[1]{%
	\vcenter{%
		\Let@ \restore@math@cr \default@tag
		\baselineskip\fontdimen10 \scriptfont\tw@
		\advance\baselineskip\fontdimen12 \scriptfont\tw@
		\lineskip\thr@@\fontdimen8 \scriptfont\thr@@
		\lineskiplimit\lineskip
		\ialign{\hfil$\m@th\scriptstyle##$&$\m@th\scriptstyle{}##$\crcr
			#1\crcr
		}%
	}
}

\title{\LARGE \bf
	Performance bounds of adaptive MPC with bounded parameter uncertainties
}

\author{Francisco Moreno-Mora, Lukas Beckenbach and Stefan Streif%
	\thanks{Francisco Moreno-Mora, Lukas Beckenbach and Stefan Streif are with the Laboratory for Automatic
		Control and System Dynamics, Technische Universität Chemnitz, 09107 Chemnitz, Germany. This work was partially sponsored by the Federal Ministry of Education and Research of Germany (Project: SOPRANN).}}

\begin{document}
	
	\SETCR{\CRELSE{https://doi.org/10.1016/j.ejcon.2022.100688}{European Journal of Control}{notused}}
	
	\maketitle
	\thispagestyle{empty}
	\pagestyle{empty}

	\begin{abstract}
		
		Model predictive control is a control approach that minimizes a stage cost over a predicted system trajectory based on a model of the system and is capable of handling state and input constraints.
		For uncertain models, robust or adaptive methods can be used.
		Because the system model is used to calculate the control law, the closed-loop behavior of the system and thus its performance, measured by the sum of the stage costs, are related to the model used.
		If it is adapted online, a performance bound is difficult to obtain and thus the impact of model adaptation is mostly unknown.
		This work provides a (worst-case) performance bound for a linear adaptive predictive control scheme with a specific model parameter estimation. 
		The proposed bound is expressed in terms of quantities such as the initial system parameter error and the constraint set, among others and can be calculated a priori. 
		The results are discussed in a numerical example.
	\end{abstract}

	\section{Introduction}
	Model predictive control (MPC) has become a standard control approach in industrial practice. 
	It iteratively solves an optimization problem, which minimizes a finite horizon sum of the so-called stage cost along the predicted trajectory of the system, calculated using a model of the system. 
	Input and state constraints, which arise frequently in applications, can be directly considered in the optimization. Because of its predictive nature, the behavior of the closed-loop system is directly related with the model used by the MPC scheme, \cite{abhijit2009modelmismatch,lu2021robustMPC}. 
	In the presence of model uncertainty various settings of MPC have been proposed that are tailored to this case \eg robust MPC \cite{bemporad1999robust,mayne2005robust,kouvaritakis2016MPC,streif2014robustNMPC}; adaptive MPC \cite{kim2008adaptive,heirung2013adaptive,tanaskovic2014adaptive}; or stochastic MPC \cite{kouvaritakis2016MPC,heirung2017dual,mesbah2014stochasticNMPC}.
	
	Most studies of MPC schemes focus only on stability and recursive feasibility. 
	Usually, so-called terminal conditions, i.e. terminal cost and terminal constraints, are used to guarantee these properties, see \cite{mayne2000constrained} for a survey.
	Since terminal conditions can be difficult to design and pose an additional burden on the optimization, schemes without such constraints have been analyzed, see \eg \cite{grune2009analysis}, \cite{grune2010analysis} and \cite{boccia2014stability}. In this case a sufficiently long optimization horizon can guarantee these properties. 
	The stability analysis of these approaches is commonly based on relaxed dynamic programming, \cite{lincoln2006relaxing,rantzer2006relaxed}. 
	Not only can stability be shown with this approach, but bounds on the infinite horizon (IH) performance under the given stage cost can also be provided, e.g. \cite{grune2009analysis}. These results have been extended in a robust setting to models with uncertainties or disturbances in \cite{schwenkel2020robust}.
	
	Adaptive MPC schemes can also be used in the case of uncertain models. These schemes use a system model that is updated online. Studies of stability and feasibility usually incorporate properties of the specific estimation algorithm used, as done in, for example, \cite{dicairano2016indirect,lorenzen2019robust,aswani2013provably}.
	Performance studies for adaptive MPC schemes are scarce.
	Some effort has lately been put into this direction in \cite{lorenzen2019robust}, where the authors provide a state norm bound for linear systems with affine uncertainties and disturbances. 
	However, the relation of the controller performance to the optimal IH cost along the true dynamics has not been addressed, even though MPC is an approximation of the solution of the infinite horizon optimal control problem.
	This work analyzes the performance of linear adaptive MPC under a specific parameter estimation approach and provides a worst-case a priori bound on the IH cost depending on particular aspects of the estimation algorithm. 
	After the control and estimation framework is established in Sec. \ref{sec:problem_setup}, the main results of Sec. \ref{sec:performance-bound} provide a computable bound on the IH cost. 
	Specifically, the optimal finite horizon (FH) cost of the MPC scheme is related to the optimal IH cost, and afterwards, the closed-loop IH cost under the applied control actions resulting from the MPC scheme with estimated parameters is incorporated.
	\subsubsection*{Notation} The $i$-th entry of a vector is denoted $[a]_i$. A column vector of ones is denoted by $\bs{1}$, a vector of zeros is denoted analogously and the identity matrix is denoted by $I$. The size of these vectors and matrices is inferred from context. Positive (semi)definite matrices are denoted $A \succ 0$ ($A \succeq 0 $). The Euclidean norm of vector $x$ is denoted by $\| x \|$ and for $S \succ 0$ define $\| x \|^2_S = x^T S x$. For a matrix $A$, $\| A \|$ denotes the induced norm of the Euclidean vector norm, i.e., $\| A\|= \sup \{\| Ax \| : x\in \R^n,\|x \|=1\}$. The set of natural numbers is denoted $\N$ and the set of natural numbers and zero is $\N_0$. $\R_{\geq 0}$ is the set of non-negative real numbers, $\R_{> 0}$ is defined analogously. $A \oplus B$ denotes the Minkowski set addition.
	
	\section{Problem setup and preliminaries}
	\label{sec:problem_setup}
	We consider discrete-time linear time-invariant systems of the form
	\begin{equation}
		\label{eq:sys} %
		x_{k+1} = A(\theta^*)x_k + B(\theta^*)u_k,
	\end{equation}
	where $x_k\in \R^n$ is the state of the system starting at $x_0$, $u_k \in \R^m$ is the control input and $\theta^* \in \R^p$ is an unknown constant parameter vector. The following assumption specifies the dependence of the system matrices on the unknown parameter vector.
	\pagebreak
	\begin{asm}[Uncertainty] \label{asm:matrices-param-affine}
		The system matrices depend affinely on the parameter vector $\theta^* \in \R^p$ \ie 
		\begin{equation}
			\label{eq:sys-matrices}
			(A(\theta^*),B(\theta^*)) = (A_0,B_0) + \sum_{i=1}^p (A_i,B_i)[\theta^*]_i,
		\end{equation} 
		which is contained in a known set $\Theta$ \ie
		\begin{equation}
			\theta^\ast \in \Theta \coloneqq \{\theta \in \R^p | H_\theta \theta \leq h_\theta\}.
		\end{equation}
	\end{asm}
	
	Oftentimes it is desirable that the state and input are constrained to a user-defined set. 
	Consider the polytopic constraint set
	\begin{align}
		\label{eq:ConstraintSet}
		\mathbb{Z} = \{ (x,u) \in \R^n \times \R^m | Fx + Gu \leq \mathbf{1} \}
	\end{align}
	with given matrices $F \in \R^{c \times n}$ and $G \in \R^{c \times m}$. 
	This type of constraint set includes \eg input saturation. 
	
	The objective of this work is to investigate the closed-loop stability of the origin and the associated infinite horizon (IH) cost function of the form
	\begin{align} \label{eq:inf-cost}
		J_\infty(x_0,\theta^\ast,\{u_k\}) \coloneqq \sum_{k=0}^{\infty} \lVert x_k \rVert_Q^2 + \lVert u_k \rVert_R^2
	\end{align}
	along system \eqref{eq:sys} under a sequence of controls $\{u_k\}_{k=0}^{\infty}$ generated by a particular MPC and parameter estimation algorithm, for any $x_0$ such that the MPC optimal control problem is initially feasible.
	Here, $Q  \in \R^{n \times n}$ and $R \in \R^{m \times m}$ satisfying $Q,R\succ 0$ are user-defined state and input weighting matrices.
	Let $\{u_k^\ast(x_0)\}_{k =0}^{\infty}$ denote the minimizing sequence to $J_{\infty}$ and $V_{\infty}(x_0) \coloneqq J_\infty(x_0,\theta^\ast,\{u_k^\ast\})$, for all $x_0$.
	
	The remainder of this section introduces the control setting of \cite{lorenzen2019robust}, whose performance is evaluated in the subsequent Sec. \ref{sec:performance-bound}.
	
	\subsection{Parameter estimation}
	First, the parameter and parameter set estimation is discussed. Denote $\Theta_k$ the so-called membership set which refers to a set of uncertain parameters and for which it holds that $\theta^\ast \in \Theta_k$, for all $k \in \N_0$.
	Let $D(x,u)\in \R ^{n \times p}$ be defined as
	\begin{equation*}
		D(x,u) \coloneqq \left[ A_1x+B_1u, \; A_2x+B_2u,\; \dotsc \;,\; A_p x + B_p u\right],
	\end{equation*}
	and 
	\begin{equation}
		\Delta_k := \{\theta \in \R^p | x_k - (A(\theta)x_{k-1} +B(\theta)u_{k-1}) = 0\}.
	\end{equation}
	Starting at some initial guess $\Theta_0 = \Theta$, let
	\begin{equation}
		\label{eq:parameter_set_update}
		\Theta_k = \Theta_{k-1} \cap \Delta_k, \; k \in \N.
	\end{equation} 
	For a particular realization of this update, the reader may consult \cite[Sec. 3]{lorenzen2019robust}.
	
	In addition to the membership set, a point estimate $\hat{\theta}_k \in \R^p$, $k \in \N_0$, of the unknown parameter is used. 
	Given $\hat{\theta}_k$ and any $\bs{u}=\{u_l\}_{l =0}^{N-1}$, $u_l \in \R^m$, denote 
	\begin{align}
		\label{eq:predicted_state}
		\begin{split}
			\hat{x}_{l+1|k}(\bs{u};x_k) &=  A(\hat{\theta}_k)\hat{x}_{l|k}(\bs{u};x_k) + B(\hat{\theta}_k)u_{l},\\ 
			\hat{x}_{0|k}(\bs{u};x_k) &=x_k
		\end{split}
	\end{align}
	for any $x_k \in \R^n$ as well as
	\begin{align}
		\label{eq:predicted_state-opt}
		\begin{split}
			x_{l+1|k}^\ast(\bs{u};x_k) &=  A(\theta^\ast)x_{l|k}^\ast(\bs{u};x_k) + B(\theta^\ast)u_{l},\\
			x_{0|k}^\ast(\bs{u};x_k) &=x_k,
		\end{split}
	\end{align}
	with which the prediction error can be defined as 
	\begin{align}
		\label{eq:prediction_error}
		\tilde{x}_{l|k}(\bs{u}) \coloneqq\, &x_{l|k}^\ast(\bs{u};x_k)-\hat{x}_{l|k}(\bs{u};x_k)   \\ 
		= &\sum_{i=0}^{l-1} A^i(\hat{\theta}_k)D(x^*_{l-1-i|k}(\bs{u};x_k),u_{l-1-i})(\theta^*-\hat{\theta}_k)  \notag
	\end{align}
	for $l \in \{1,\dotsc,N\}$, noting that $\tilde{x}_{0|k}(\bs{u}) = 0$. 
	Starting at a known initial guess $\hat{\theta}_0 \in \Theta$, the point estimate $\hat{\theta}_k$ is obtained via
	\begin{align}
		\begin{split}
			\label{eq:point_estimate_update}
			\hat{\theta}_k = \prod_\Theta (\theta_k^{\text{est}}&),\\
			\theta^{\text{est}}_k := \hat{\theta}_{k-1} &+ \\  \mu& D(x_{k-1},u_{k-1})^{\green{\top}}(x_k-\hat{x}_{1|k-1}(\bs{u}_{k-1};x_{k-1})) 
		\end{split}
	\end{align}
	where $\mu \in \R_{>0}$ denotes the gain satisfying $\frac{1}{\mu}>\sup_{(x,u)\in \mathbb{Z}} \lVert D(x,u) \rVert^2$ and $\prod_\Theta(\hat{\theta})= \argmin_{\theta \in \Theta} \lVert \theta - \hat{\theta} \rVert$ denotes the Euclidean projection of $\hat{\theta} \in \R^p$ onto the set $\Theta$. Let $\tilde{\theta}_k:=\theta^*-\hat{\theta}_k$.
	
	The following result will be used in the subsequent performance study:
	\begin{lem}[\hspace*{-3pt}\cite{lorenzen2019robust}] 
		\label{lem:ParamEstimation} 
		If $\sup_{k\in \N} \lVert x_k \rVert<\infty$, $\sup_{k\in \N} \lVert u_k \rVert<\infty$, then $\hat{\theta}_k \in \Theta$ for all $k \in \N$ along \eqref{eq:point_estimate_update}, with $\hat{\theta}_0 \in \Theta$, and
		\begin{equation}
			\label{eq:lemma-1}
			\sup_{m \in \N,\hat{\theta} \in \Theta} \frac{\sum_{k=0}^m \lVert \tilde{x}_{1|k} \rVert^2}{\frac{1}{\mu}\lVert \theta^\ast - \hat{\theta} \rVert^2 }\leq 1.
		\end{equation}
	\end{lem}
	\vspace{1em}
	From eq. \eqref{eq:prediction_error}, using the bound $\mu$ and the sum of a geometric progression
	\begin{align}
		\label{eq:bound_prediction_error}
		\lVert \tilde{x}_{l|k}(\bs{u}) \rVert^2 \leq \frac{c_1^2(l,\hat{\theta}_k)}{\mu}\lVert \theta^\ast - \hat{\theta}_k \rVert^2
	\end{align}
	in which
	\begin{align*}
		c_1(l,\hat{\theta}_k) \coloneqq \begin{dcases}\frac{1-\lVert A(\hat{\theta}_k) \rVert^l}{1-\lVert  A(\hat{\theta}_k) \rVert} &\lVert A(\hat{\theta}_k) \rVert \neq 1 \\
			l &\text{else}.
		\end{dcases}
	\end{align*}
	\begin{rem}
		Lemma \ref{lem:ParamEstimation} is a consequence of the dynamics of the chosen parameter estimation algorithm. Using similar prediction error bounds to \eqref{eq:lemma-1} for other algorithm choices based on properties such as, for example, convergence speed, the subsequent performance analysis could be extended to other estimation schemes. This also holds for algorithms which use noisy measurements, which are  beyond the scope of this work.
	\end{rem}
	
	The utilized MPC scheme uses the point estimate and the membership set as described in the following section.
	
	\subsection{Adaptive MPC}
	
	The control sequence applied to the system is obtained via a tube-based adaptive MPC. 
	The control action is determined using the parametrization
	\begin{equation} \label{eq:input-parametrization}
		u_{l|k}(x) = Kx + v_{l|k},
	\end{equation}
	where $\bs{v}_{k} \coloneqq \{ v_{l|k} \}_{l =0}^{N-1}$ denotes MPC decision variables and $K \in \R^{m \times n}$ is a feedback gain that satisfies the following assumption
	\begin{asm} 
		\label{asm:stabilizing-k}
		For all $\theta \in \Theta$, $A_{cl}(\theta) \coloneqq A(\theta) + B(\theta)K$ is stable. 
	\end{asm}
	\begin{rem}
		\label{rem:stabilizing-k}
		It may not be possible to find $K$ such that assumption \ref{asm:stabilizing-k} is fulfilled, for example when $(A(\theta),B(\theta))$ is not stabilizable for some $\theta \in \Theta$. If such a $K$ exists, it can be determined using linear matrix inequalities using the method in \cite{deoliveira1999robust}.
	\end{rem}
	
	Consider an auxiliary state tube sequence $\{\mathbb{X}_{l|k} \}_{l = 0}^{N}$ which satisfies
	\begin{equation}
		\label{eq:state_tube}
		\begin{gathered}
			\mathbb{X}_{0|k} \ni x_k \\
			\mathbb{X}_{l+k|k} \ni A(\theta)x+B(\theta)u_{l|k}(x)\quad \forall x \in \mathbb{X}_{l|k}, \theta \in \Theta_k, \\
			x \times u_{l|k}(x) \in \mathbb{Z} \quad \forall x \in \mathbb{X}_{l|k}.
		\end{gathered}
	\end{equation}
	
	The sets $\mathbb{X}_{l|k}$ are recursively calculated as
	\begin{equation}
		\mathbb{X}_{l|k} = \{z_{l|k}\} \oplus \alpha_{l|k} \mathbb{X}_0
	\end{equation}
	with a set of optimization variables $\bs{z}_k \coloneqq \{z_{l|k}\}_{l=0}^{N}$, $z_{l|k} \in \R^n$, and $\bs{\alpha}_k \coloneqq \{\alpha_{l|k}\}_{l=0}^{N}$, $\alpha_{l|k} \in \R_{\geq 0}$, such that \eqref{eq:state_tube} holds, based on a given polytope $\mathbb{X}_0$.
	
	Consider further the following
	\begin{asm}
		\label{asm:terminal-set}
		There exist a nonempty set $\mathbb{X}_f \coloneqq \{ (z,\alpha) \in \R^n \times \R_{\geq 0} | H_Tz+h_T\alpha \leq 1\}$ such that $(x,Kx)\in \mathbb{Z}$ for all $x \in \{z\} \oplus \alpha \mathbb{X}_0,(z,\alpha)\in \mathbb{X}_f$ and for all $\theta \in \Theta$,
		\begin{align*}
			(z,\alpha) \in \mathbb{X}_f \implies &\exists(z^+,\alpha^+) \in \mathbb{X}_f \quad \text{s.t.} \\
			&A_{cl}(\theta)(\{z\} \oplus \alpha \mathbb{X}_0) \subseteq \{ z^+ \} \oplus \alpha^+ \mathbb{X}_0.
		\end{align*}
		
	\end{asm}
	Define the cost function
	\begin{equation}
		\label{eq:MPC_cost}
		\begin{split}
			&J_N(x_k,\hat{\theta}_k,\bs{v}_{k}) \\
			\coloneqq &\sum_{l=0}^{N-1} \lVert \hat{x}_{l|k}(\bs{u}_k;x_k) \rVert_Q^2 + \lVert u_{l|k} \rVert_R^2 + \lVert \hat{x}_{N|k}(\bs{u}_k;x_k) \rVert_P^2
		\end{split}
	\end{equation}
	with \eqref{eq:predicted_state} using $\hat{\theta}_k$ and $u_{l|k}$ as defined in \eqref{eq:input-parametrization}, where $P \in \R^{n \times n},P \succ 0$ is such that
	\begin{equation}
		\label{eq:matrix_P_cond}
		A_{cl}(\theta)^T P A_{cl}(\theta) + Q + K^T R K \preccurlyeq P \quad \forall \theta \in \Theta.
	\end{equation}
	\begin{rem}
		\label{rem:P-for-all-theta}
		A matrix $P$ that fulfills \eqref{eq:matrix_P_cond} can be found in a similar manner as in \cite[Thm. 1]{deoliveira1999robust}.
	\end{rem}
	
	Henceforth, all decision variables are grouped in $\bs{d}_{k}:= \{ \bs{z}_{k}, \bs{\alpha}_{k}, \bs{v}_{k} \}$. 
	For any $(x_k,\Theta_k)$, let 
	\begin{align*}
		\mathbb{D}(x_k,\Theta_k) \coloneqq \{ \bs{d}_{k} \lvert \eqref{eq:state_tube}, (z_{N|k},\alpha_{N|k}) \in \mathbb{X}_f  \}
	\end{align*}
	be the set of admissible decision variables.

	At each time step $k$, the predictive control and parameter estimation scheme finds a minimizing collection of sequences $\bar{\bs{d}_k} \coloneqq \{ \bar{\bs{z}}_{k}, \bar{\bs{\alpha}}_{k}, \bar{\bs{v}}_{k} \}$ to
	\begin{equation}
		\label{eq:mpc_estimate}
		\begin{gathered}
			\min_{\bs{d}_{k}} \quad J_N(x_k,\hat{\theta}_k,\bs{v}_{k})\\
			\text{s.t.}\quad \bs{d}_{k} \in \mathbb{D}(x_k,\Theta_k)
		\end{gathered}
	\end{equation}
	based on the current estimate $\hat{\theta}_k$, $\Theta_k$ and measurement $x_k$. 
	To comply with \cite{lorenzen2019robust}, denote the minimum to \eqref{eq:mpc_estimate} by $V_N(x_k,\hat{\theta}_k,\Theta_k) \coloneqq J_N(x_k,\hat{\theta}_k,\bar{\bs{v}}_{k})$. 
	For convenience, let $\bar{\bs{u}}_k \coloneqq \{\bar{u}_{l|k}\}$, $\bar{u}_{l|k} = K \hat{x}_{l|k}(\bar{\bs{u}}_k;x_k) + \bar{v}_{l|k}$, of which the first element $u_k^{\text{MPC}} \coloneqq u_{0|k}$ is applied to the system and where $\hat{x}_{l|k}(\bar{\bs{u}}_k;x_k)$ is the optimal state sequence under $\bar{\bs{u}}_k$. It should be noted that the scheme is guaranteed to be recursively feasible if it is initially feasible.
	
	\section{Closed-loop IH performance bound}
	\label{sec:performance-bound}
	
	This section deals with the estimation of the closed-loop performance under the proposed adaptive MPC scheme. Due to the adaptive nature of the prediction model, obtaining an a priori performance bound is difficult as adaptation depends on the true trajectory taken by system. 
	However, a performance bound can be provided with particular consideration of the update gain $\mu$ as well as the parameter error $\tilde{\theta}_k$ at $k=0$.
	
	This section is split into two parts: first, a relation between the cost $V_N$ and $V_{\infty}$ is established. 
	Then, the closed-loop cost under $\{u_k^{\text{MPC}}\}$ is inspected and a bound thereof provided.
	
	\begin{prp} \label{prop:infinite_vs_MPC_VF}
		Let Asm. \ref{asm:matrices-param-affine}-\ref{asm:terminal-set} hold. 
		For any $\Z \subset \R^n \times \R^m$ in the form of \eqref{eq:ConstraintSet}, any $(x_k,\Theta_k)$ such that $\mathbb{D}(x_k,\Theta_k) \neq \emptyset$, any $\hat{\theta}_k \in \Theta_k$ and $k \in \N_0$, there exist $c_{V},c_f,\bar{\Delta}\geq 0$ and a function $d_{\tilde{\theta}}:\R_{>0} \times \R_{>0} \ra \R_{\geq 0}$ such that the following holds
		\begin{align} \label{eq:V_N-up-bnd}
			V_N(x_k,\hat{\theta}_k,\Theta_k) \leq c_V V_\infty(x_k) + d_{\tilde{\theta}}(\lVert \tilde{\theta}_k \rVert,\mu) + \bar{\Delta} + c_f.
		\end{align}
	\end{prp}
	
	\begin{proof}
		By definition,
		\begin{align*}
			V_N(&x_k,\hat{\theta}_k,\Theta_k) \\
			= &\sum_{l=0}^{N-1}  \lVert \hat{x}_{l|k}(\bs{\bar{u}}_k;x_k) \rVert_Q^2 + \lVert \bar{u}_{l|k} \rVert_R^2 + \lVert \hat{x}_{N|k}(\bs{\bar{u}}_k;x_k) \rVert_{P} ^2.
		\end{align*}
		Note that by the Cauchy-Schwarz and Young's inequalities, it holds that for any $y,z\in \R^n$, $S \succ 0$ and $\varepsilon \in \R_{>0}$, 
		\begin{equation}
			\label{eq:young_cauchyschwarz}
			\lVert y-z \rVert_S^2 \leq (1+\varepsilon) \lVert y \rVert^2_S + \left( 1+\frac{1}{\varepsilon} \right) \lVert z \rVert^2_S .
		\end{equation}
		
		Let $c_f \coloneqq \max_{x\in \{z_{N|k}\} \oplus \alpha_{N|k} } \|x\|_P^2$ as well as $c_Q \coloneqq \lVert Q \rVert$. 
		Employing \eqref{eq:young_cauchyschwarz} on $\hat{x}_{l|k}(\bs{u};x_k) = x_{l|k}^\ast(\bs{u};x_k)- \tilde{x}_{l|k}(\bs{u})$ from \eqref{eq:prediction_error} and further using \eqref{eq:bound_prediction_error}, it follows that
		\begin{equation} \label{eq:V_N-bound-true-param}
			\begin{split} 
				V_N(x_k,\hat{\theta}_k,\Theta_k)\leq  &\sum_{l=0}^{N-1} \Biggl( (1+\varepsilon_1)\lVert x^*_{l|k}(\bar{\bs{u}}_k;x_k) \rVert_Q^2 +  \lVert \bar{u}_{l|k} \rVert_R^2 \\ 
				&+  \frac{c_Q c_1(l,\hat{\theta}_k)^2}{\mu^2}\left(1+\frac{1}{\varepsilon_1}\right) \lVert \tilde{\theta}_k \rVert^2 \Biggr) + c_f
			\end{split}
		\end{equation}
		for all $\eps_1 \in \R_{>0}$, making use of the fact that $\hat{x}_{N|k}(\bs{\bar{u}}_k;x_k)\in \{z_{N|k}\} \oplus \alpha_{N|k} \mathbb{X}_0$, $(z_{N|k},\alpha_{N|k}) \in  \mathbb{X}_f$. 
		
		For any $k \in \N_0$, define $ \delta u_{l|k} \coloneqq u_l^\ast(x_k) - \bar{u}_{l|k}$, $l \in \{0,\dotsc,N-1\}$ and recall that 
		\begin{equation*}
			x^*_{l|k}(\bs{u}_k+\bs{w}_k;x_k)=x^*_{l|k}(\bs{u}_k;x_k)+x^*_{l|k}(\bs{w}_k;\bs{0})
		\end{equation*}
		for any $\bs{u}_k$, $\bs{w}_k$. 
		Applying \eqref{eq:young_cauchyschwarz} to $\bar{u}_{l|k}$ and $x^*_{l|k}(\bar{\bs{u}}_k;x_k)$, both times with the same $\varepsilon = \varepsilon_2 \in \R_{>0}$, it follows that
		\begin{align*}
			\sum_{l=0}^{N-1} (1+&\varepsilon_1)\lVert x^*_{l|k}(\bar{\bs{u}}_k;x_k) \rVert^2_Q +  \lVert \bar{u}_{l|k} \rVert_R^2 \\
			\leq & \sum_{l=0}^{N-1} \Biggl( (1+\varepsilon_1)(1+\varepsilon_2)\lVert x^*_{l|k}(\bs{u}^\ast_k;x_k) \rVert^2_Q 
		\end{align*}
		
		\begin{align}
			\label{eq:V_N_bound_intermediate_res}
			\begin{split}
				&+(1+\varepsilon_1)(1+\varepsilon_2)\lVert u^*_{l}(x_k) \rVert^2_R  \\
				&+ (1+\varepsilon_1)\left(1+\frac{1}{\varepsilon_2}\right) \lVert  x^*_{l|k}(\bs{\delta u}_k;\bs{0}) \rVert^2_Q \\
				&+ \left(1+\frac{1}{\varepsilon_2} \right)\lVert \delta u_{l|k} \rVert^2_R \Biggr)
			\end{split}
		\end{align}
		in which $\| u^*_{l}(x_k) \|_R^2$ is additionally magnified by $(1+\varepsilon_1)$ with $\varepsilon_1 \in \R_{>0}$.
		Recall that
		\begin{align*}
			x^*_{l|k}(\bs{\delta u};\bs{0}) = \sum_{i=0}^{l-1}A(\theta^*)^i B(\theta^*) \delta u_{l-i-1|k} 
		\end{align*}
		for $l \in \{1,\dotsc,N\}$ is the closed form for the state under $\delta u$ starting at zero initial state.
		Denoting $c_B = \max_{\theta \in \Theta_k} \lVert B(\theta) \rVert^2$, $c_{A} = \max_{\theta \in \Theta_k} \lVert A(\theta) \rVert^2$ and $c_R = \lVert R \rVert$, the above inequality can further be written as
		\begin{align*}
			\sum_{l=0}^{N-1} &(1+\varepsilon_1)\lVert x^*_{l|k}(\bar{\bs{u}}_k;x_k) \rVert^2_Q +  \lVert \bar{u}_{l|k} \rVert_R^2 \\
			\leq &(1+\varepsilon_1)\left(1+\varepsilon_2\right)V_\infty(x_k) \\ 
			&+ \sum_{l=0}^{N-1}\Biggl( (1+\varepsilon_1)\left(1+\frac{1}{\varepsilon_2}\right) c_Qc_B \sum_{i=0}^{l-1}c_A^{i} \lVert \delta u_{l-i-1|k}\rVert^2 \\
			&+\left(1+\frac{1}{\varepsilon_2} \right) c_R  \lVert \delta u_{l|k} \rVert^2 \Biggr)
		\end{align*}
		where $V_{\infty}(x_k)$ has been employed as an upper bound on the finite horizon cost.
		
		It is possible to bound $\|\delta u_{l|k}\|^2$ for any given tuple $\{x_k$, $\hat{\theta}_k$, $\Theta_k\}$, as shown by an example in the appendix. 
		It should be noted that the presented bound is associated with the constraint set $\Z$, the parameter estimate $\hat{\theta}_k$ and the membership set $\Theta_k$ and may be improved upon, which however is beyond the scope of this work.
		Denote a particular bound by $\|\delta \bar{u}^*_{l|k}\|^2 \geq \|\delta u_{l|k}\|^2$. 
		Then, it follows from \eqref{eq:V_N-bound-true-param} that
		\begin{equation}
		\label{eq:prop1-final-ineq}
		\begin{split}
			V_N(x_k,\hat{\theta}_k,&\Theta_k) \leq 
			\underbrace{(1+\varepsilon_1)\left(1+\varepsilon_2\right)}_{c_V}V_\infty(x_k) \\
			&+\underbrace{(1+\varepsilon_1)\left(1+\frac{1}{\varepsilon_2}\right) c_Qc_B \sum_{l=0}^{N-1}\sum_{i=0}^{l-1}c_A^{i} \|\delta \bar{u}^*_{l|k}\|^2}_{  \eqqcolon \bar{\Delta}_1} \\
			&+\underbrace{\left(1+\frac{1}{\varepsilon_2} \right) c_R  \sum_{l=0}^{N-1}\|\delta \bar{u}^*_{l|k}\|^2}_{ \eqqcolon \bar{\Delta}_2} \\
			&+ \underbrace{\sum_{l=0}^{N-1}\frac{c_Q c_1(l,\hat{\theta}_k)^2}{\mu^2}\left(1+\frac{1}{\varepsilon_1}\right) \lVert\tilde{\theta}_k \rVert^2}_{ \eqqcolon d_{\tilde{\theta}}(\lVert\tilde{\theta}_k \rVert,\mu)} + c_f,
		\end{split}
		\end{equation}
		and with $\bar{\Delta}=\bar{\Delta}_1+\bar{\Delta}_2$, inequality \eqref{eq:V_N-up-bnd} is obtained.
	\end{proof}
	
	The next theorem is our main result and relates the infinite horizon closed-loop cost of system \eqref{eq:sys} under the adaptive MPC \ie $J_{\infty}(x_0,\theta^\ast,\{u_k^{\text{MPC}}\}) \eqqcolon J_{\infty}^{\text{MPC}}(x_0)$, with the value function of the IH control problem $V_\infty(x_0)$.
	
	\begin{thm} 
		\label{thm:performance_bound}
		Let Asm. \ref{asm:matrices-param-affine}-\ref{asm:terminal-set} hold. 
		For any $\mathbb{Z} \subseteq \R^n \times \R^m$, any $(x_0,\Theta_0)$ such that $\mathbb{D}(x_0,\Theta_0) \neq \emptyset$ and any $\hat{\theta}_0 \in \Theta_0$, there exist $\alpha_{V},\alpha_{f},\alpha_{\Delta} \geq 0$ and a function $a:\R_{>0} \times \R_{>0} \ra \R_{\geq 0}$ such the following bound holds
		\begin{align*}
			J_{\infty}^{\text{MPC}}(x_0)  \leq  \alpha_V V_\infty(x_0)  + \alpha_f + \alpha_{\Delta}  + a(\|\tilde{\theta}_0\|,\mu)
		\end{align*}
		along the closed-loop \eqref{eq:sys} under $\{u_k^{\text{MPC}}\}$, point estimate \eqref{eq:point_estimate_update} and the membership set update \eqref{eq:parameter_set_update}.
	\end{thm}
	
	\begin{proof}
		Recall from \cite[Thm. 14]{lorenzen2019robust} that the sequence $v_{l|k+1} = \bar{v}_{l+1|k}$ for $l\in \{ 0,\dots,N-2 \}$ and $v_{N-1|k+1}=0$ is feasible and that for $\delta \hat{x}_{l|k} \coloneqq \hat{x}_{l-1|k+1}(\bs{u}_{k+1};x_{k+1})-\hat{x}_{l|k}(\bs{\bar{u}}_k;x_k)$, $u_{k+1}=K\hat{x}_{l-1|k+1}(\bs{u}_{k+1};x_{k+1}) +v_{l|k+1}$, it holds that
		\begin{align*}
			\lVert \delta \hat{x}_{l|k} \rVert \leq \left( \sum_{i=0}^{l-1} \lVert A_{cl}(\hat{\theta}_{k+1}) \rVert^{l-i} \right) \lVert \tilde{x}_{1|k}(\bs{\bar{u}}_k) \rVert.
		\end{align*}
		Therefore, for any $S \succ 0$, any $k \in \N_0$ and $l \in \{1,\dotsc,N\}$,
		\begin{align}
			\label{eq:deltaX_bound}
			\lVert \delta \hat{x}_{l|k} \rVert^2_S \leq \lVert S \rVert \lVert \delta \hat{x}_{l|k} \rVert^2 \leq c_2(l) \lVert S \rVert \lVert \tilde{x}_{1|k}(\bs{\bar{u}}_k) \rVert^2
		\end{align}
		where 
		\begin{equation*}
			c_2(l) \coloneqq \begin{dcases}\frac{1-c_{cl}^{l+1}}{1-c_{cl}}-1 & c_{cl} \neq 1\\
				l &\text{else}.
			\end{dcases}
		\end{equation*}
		with $c_{cl} = \max_{\theta \in \Theta} \| A_{cl}(\theta) \|^2$. 
		
		Consider the following the difference
		\begin{align*}
			V_N(x_{k+1},&\hat{\theta}_{k+1},\Theta_{k+1}) - V_N(x_k,\hat{\theta}_k,\Theta_k)  \\ 
			\leq &-\lVert x_k \rVert_Q^2 - \lVert u_k^{\text{MPC}} \rVert_R^2 + \varepsilon_3 V_N (x_k,\hat{\theta}_k,\Theta_k)\\
			&+\left(1+\frac{1}{\varepsilon_3}\right) \left(\sum_{l=1}^{N-1}\lVert \delta \hat{x}_{l|k} \rVert_{\overline{Q}}^2 + \lVert \delta \hat{x}_{N|k} \rVert_P^2 \right) \\
		\end{align*}
		which has been observed in \cite{lorenzen2019robust}, with $\overline{Q}=Q+K^\top R K$, for any $\varepsilon_3 \in \R_{>0}$.  Denoting
		\begin{equation*}
			c_3 \coloneqq  \sum_{l=1}^{N-1} c_2(l) \lVert \overline{Q} \rVert + c_2(N) \lVert P \rVert
		\end{equation*}
		the cost difference can be further bounded by
		\begin{align*}
			V_N(x_{k+1}&,\hat{\theta}_{k+1},\Theta_{k+1}) - V_N(x_k,\hat{\theta}_k,\Theta_k)\\
			\leq  &-\lVert x_k \rVert_Q^2 - \lVert u_k^{\text{MPC}} \rVert_R^2 + \varepsilon_3 c_\theta \lVert x_k \rVert^2 \\
			&+ \left( 1 + \frac{1}{\varepsilon_3} \right) \sum_{l=1}^{N-1} c_2(l) \lVert \overline{Q} \rVert \lVert \tilde{x}_{1|k}(\bs{\bar{u}}_k) \rVert^2  \\
			&+ \left(1 + \frac{1}{\varepsilon_3} \right) c_2(N) \lVert P \rVert \lVert \tilde{x}_{1|k}(\bs{\bar{u}}_k) \rVert^2 \\
			\leq &-\lVert x_k \rVert^2_Q - \lVert u_k^{\text{MPC}} \rVert^2_R  + \left( 1 + \frac{1}{\varepsilon_3} \right)c_3 \lVert \tilde{x}_{1|k}(\bs{\bar{u}}_k) \rVert^2\\
			&+\varepsilon_3 c_\theta \lVert x_k\rVert^2 ,
		\end{align*}
		where $c_{\theta} \|x\|^2$ is a computable bound on the finite horizon cost $V_N$ (see \eg \cite[Thm. 14]{lorenzen2019robust} or \cite{bemporad2002explicit}). 
		
		Then, adding the term $\varepsilon_3 c_{\theta} \|u_k^{\text{MPC}} \|^2$, it can be observed that
		\begin{align*}
			&-\lVert x_k \rVert^2_Q - \lVert u_k^{\text{MPC}} \rVert^2_R +\varepsilon_3 c_\theta \lVert x_k\rVert^2 + \left( 1 + \frac{1}{\varepsilon_3} \right)c_3 \lVert \tilde{x}_{1|k}(\bs{\bar{u}}_k) \rVert^2  \\
			&\leq -\gamma \left( \lVert x_k \rVert^2_Q + \lVert u_k^{\text{MPC}} \rVert^2_R\right) + \left( 1+\frac{1}{\varepsilon_3} \right)c_3 \lVert \tilde{x}_{1|k}(\bs{\bar{u}}_k) \rVert^2,
		\end{align*}
		
		where $\gamma \coloneqq (1 - \varepsilon_3 c_\theta\lVert \tilde{Q} \rVert^2)$ in which $\tilde{Q}$ is a matrix such that $\tilde{Q}^T Q \tilde{Q}=I$. 
		
		Summing both sides of the previous inequality from $0$ to $m$ and using Lem.\ref{lem:ParamEstimation} yields	
		\begin{align*}
			\gamma \; \sum_{k=0}^{m} \lVert x_k \rVert^2_Q &+ \lVert u_k^{\text{MPC}} \rVert^2_R \\
			\leq  &V_N(x_0,\hat{\theta}_0,\Theta_0) + \left( 1+ \frac{1}{\varepsilon_3} \right)  \frac{c_3}{\mu} \lVert \tilde{\theta}_0 \rVert^2.
		\end{align*}
		
		Choosing $\varepsilon_3$ s.t. $\gamma > 0$, letting $m\to \infty$ and using proposition \ref{prop:infinite_vs_MPC_VF}, we finally obtain the bound
		\begin{equation}
			\label{eq:infinite-cost-bound}
			\begin{split}
			J_\infty^{\text{MPC}}(x_0) \leq &  \underbrace{\frac{c_V}{\gamma}}_{\alpha_V} V_\infty(x_0) + \underbrace{\frac{\Delta}{\gamma}}_{\alpha_\Delta} + \underbrace{\frac{c_f}{\gamma}}_{\alpha_f} \\
			&+ \underbrace{\frac{1}{\gamma} d_{\theta}(\| \tilde{\theta}_0 \|)  +\left( 1+ \frac{1}{\varepsilon_3} \right)  \frac{c_3}{\gamma\mu} \lVert \tilde{\theta}_0 \rVert^2}_{a(\| \tilde{\theta}_0\|,\mu)}
			\end{split}
		\end{equation}
	\end{proof}
	
	This theorem provides an a priori bound for the total infinite horizon closed-loop cost in relation to the optimal infinite horizon cost, based on various available quantities, notably the initial parameter error, the constraint set and the MPC design parameters.
	
	\begin{rem}
		The proposed bound holds for any $\varepsilon_1,\varepsilon_2,\varepsilon_3 \in \R_{>0}$ such that $\gamma>0$, which are contained in the scalars seen in inequality \eqref{eq:infinite-cost-bound}. 
		Heuristically, one may compute scalars to obtain the lowest performance bound via solving
		\begin{align*}			
				\label{eq:heuristic_epsilon}	
				\min_{\varepsilon_1,\varepsilon_2,\varepsilon_3 \in \R_{>0}} \quad &\lambda\| \alpha_V -1\|^2 +\alpha_\Delta + \alpha_f + a(\lVert \tilde{\theta}_0 \rVert,\mu)\\
				\text{s.t.} \quad \quad &\gamma > 0				
		\end{align*}
		where $\lambda\in \R_{>0}$ is a weighting factor and the explicit dependence of each term on $\varepsilon_i$ is omitted.
	\end{rem}

	\section{Numerical example}
	\label{sec:simulation}
	Consider a second-order system \eqref{eq:sys} specified by matrices \eqref{eq:sys-matrices} with:
	\begin{align*}	
		A_0 &= \begin{bmatrix}
			0.9 & 0.3\\ 0 & -0.3
		\end{bmatrix}, 
		&&B_0 = \begin{bmatrix}
			0.4 \\ 0
		\end{bmatrix}, \\
		A_1 &= 0_{2\times2},
		&&B_1=\begin{bmatrix}
			0.5\\
			0
		\end{bmatrix},\\ 
		A_2& = \begin{bmatrix}
			0.143 & -0.025\\ -0.041 & 0.298
		\end{bmatrix}, && B_2 = \begin{bmatrix}
			-0.12\\-0.30
		\end{bmatrix} ,\\
		A_3 &= \begin{bmatrix}
			0.282 & 0.134\\ 0.283 & -0.242
		\end{bmatrix},
		&&B_3 = 0_{2\times 1},
	\end{align*}
	with $x_0=[-3\;-1]^\top$, $\theta^* = [0.5\;0.5\;0.75]^\top$, state constraints $|[x]_1|\leq 5$, $-5\leq [x]_2\leq 1.5$ and input constraints $|u|\leq 6$. The MPC horizon was set to $N=10$. The weighting matrices are
	\begin{align*}
		Q = \begin{bmatrix}
			1 & 0\\0 & 1
		\end{bmatrix},\quad R=1,\quad P=\begin{bmatrix}
			13.34 & 2.54\\2.54 & 2.28
		\end{bmatrix}
	\end{align*}
	and the prestabilizing feedback gain is $K=\begin{bmatrix} -1.58\;-0.57\end{bmatrix}$, where matrices $K$ and $P$ were determined as specified in Remarks \ref{rem:stabilizing-k} and \ref{rem:P-for-all-theta}. The bound provided by Thm. \ref{thm:performance_bound} depends on various quantities. In this example, the change in closed-loop behavior of the system and how it relates to the proposed performance bound, will be investigated, focusing on the membership set $\Theta_0$ and the initial parameter error $\tilde{\theta}_0$.
	
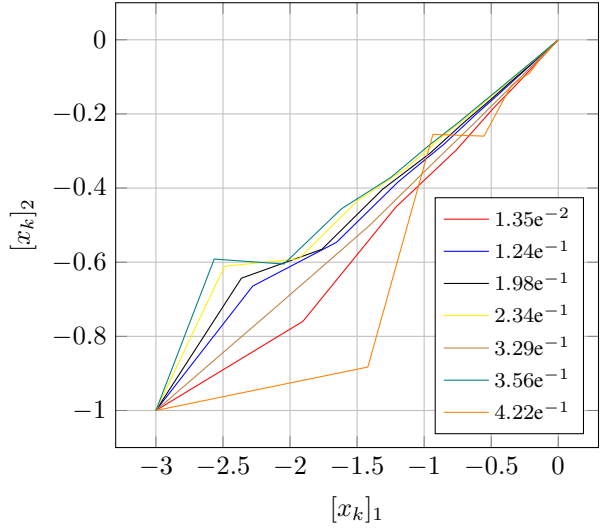
\begin{figure}
	\vspace{10pt}
	\centering
	\begin{tikzpicture}
		\begin{axis}[
			cycle list name=color list,xlabel={$[x_k]_1$}, ylabel={$[x_k]_2$}, grid=major, legend pos=south east,width=8cm,height=7.5cm,legend style={font=\footnotesize}],
			\addplot
			table[x=x,y=y,col sep=comma]{gfx/data2plot/different_big_thetas/trajectories3.csv};
			\addlegendentry{$1.35\text{e}^{-2}$};
			\addplot
			table[x=x,y=y,col sep=comma]{gfx/data2plot/different_big_thetas/trajectories7.csv};
			\addlegendentry{$1.24\text{e}^{-1}$};
			\addplot
			table[x=x,y=y,col sep=comma]{gfx/data2plot/different_big_thetas/trajectories9.csv};
			\addlegendentry{$1.98\text{e}^{-1}$};
			\addplot
			table[x=x,y=y,col sep=comma]{gfx/data2plot/different_big_thetas/trajectories10.csv};
			\addlegendentry{$2.34\text{e}^{-1}$};
			\addplot
			table[x=x,y=y,col sep=comma]{gfx/data2plot/different_big_thetas/trajectories12.csv};
			\addlegendentry{$3.29\text{e}^{-1}$};
			\addplot
			table[x=x,y=y,col sep=comma]{gfx/data2plot/different_big_thetas/trajectories13.csv};
			\addlegendentry{$3.56\text{e}^{-1}$};
			\addplot
			table[x=x,y=y,col sep=comma]{gfx/data2plot/different_big_thetas/trajectories14.csv};
			\addlegendentry{$4.22\text{e}^{-1}$};
		\end{axis}
	\end{tikzpicture}
	\caption{Closed-loop trajectory for $\Theta_0$ with different volumes ($V_{\Theta_0}$).}
	\label{fig:traj_big_thetas}
\end{figure}

\begin{figure}
	\centering
	\begin{tikzpicture}
		\begin{axis}[
			xmin=0,xmax=0.45,width=8cm,height=7.5cm,grid=major,boxplot,
			boxplot/draw direction=y,boxplot/box extend=0.012,
			cycle list={{black}},xlabel={$V_{\Theta_0}$},  ylabel={$J^{\text{MPC}}_\infty$}]
			\addplot+[boxplot/draw position=5e-4] table[y index=0, row sep=newline ]{gfx/data2plot/big_thetas_random_study/max_costs_1.csv};
			\addplot+[boxplot/draw position=4e-3] table[y index=0, row sep=newline ]{gfx/data2plot/big_thetas_random_study/max_costs_2.csv};		
			\addplot+[boxplot/draw position=0.0135] table[y index=0, row sep=newline]{gfx/data2plot/big_thetas_random_study/max_costs_3.csv};	
			\addplot+[boxplot/draw position=0.032] table[y index=0, row sep=newline ]{gfx/data2plot/big_thetas_random_study/max_costs_4.csv};                             		
			\addplot+[boxplot/draw position=0.0625] table[y index=0, row sep=newline ]{gfx/data2plot/big_thetas_random_study/max_costs_5.csv};		
			\addplot+[boxplot/draw position=0.0963] table[y index=0, row sep=newline ]{gfx/data2plot/big_thetas_random_study/max_costs_6.csv};		
			\addplot+[boxplot/draw position=0.1238] table[y index=0, row sep=newline ]{gfx/data2plot/big_thetas_random_study/max_costs_7.csv};		
			\addplot+[boxplot/draw position=0.1375] table[y index=0, row sep=newline ]{gfx/data2plot/big_thetas_random_study/max_costs_8.csv};		
			\addplot+[boxplot/draw position=0.1513] table[y index=0, row sep=newline ]{gfx/data2plot/big_thetas_random_study/max_costs_9.csv};		
			\addplot+[boxplot/draw position=0.1650] table[y index=0, row sep=newline ]{gfx/data2plot/big_thetas_random_study/max_costs_10.csv};		
			\addplot+[boxplot/draw position=0.198] table[y index=0, row sep=newline ]{gfx/data2plot/big_thetas_random_study/max_costs_11.csv};		
			\addplot+[boxplot/draw position=0.216] table[y index=0, row sep=newline ]{gfx/data2plot/big_thetas_random_study/max_costs_12.csv};		
			\addplot+[boxplot/draw position=0.234] table[y index=0, row sep=newline ]{gfx/data2plot/big_thetas_random_study/max_costs_13.csv};		
			\addplot+[boxplot/draw position=0.2535] table[y index=0, row sep=newline ]{gfx/data2plot/big_thetas_random_study/max_costs_14.csv};		
			\addplot+[boxplot/draw position=0.294] table[y index=0, row sep=newline ]{gfx/data2plot/big_thetas_random_study/max_costs_15.csv};		
			\addplot+[boxplot/draw position=0.328] table[y index=0, row sep=newline ]{gfx/data2plot/big_thetas_random_study/max_costs_16.csv};		
			\addplot+[boxplot/draw position=0.3559] table[y index=0, row sep=newline ]{gfx/data2plot/big_thetas_random_study/max_costs_17.csv};		
			\addplot+[boxplot/draw position=0.3833] table[y index=0, row sep=newline ]{gfx/data2plot/big_thetas_random_study/max_costs_18.csv};		
			\addplot+[boxplot/draw position=0.4219] table[y index=0, row sep=newline ]{gfx/data2plot/big_thetas_random_study/max_costs_19.csv};		
		\end{axis}
	\end{tikzpicture}
	\caption{Closed-loop costs for $\Theta_0$ with different volumes ($V_{\Theta_0}$).}
	\label{fig:cost_big_thetas}
	\end{figure}
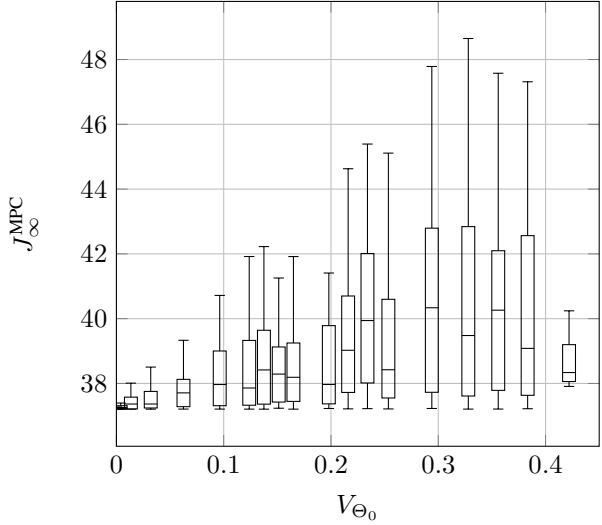

	Fig. \ref{fig:traj_big_thetas} shows the trajectory of the system under the MPC scheme for initial membership sets of the form \linebreak $\Theta_{0j}=\{ \theta \in \R^3 \;|\; a_{ij} \leq [ \theta]_i \leq b_{ij} \} $ with different sizes, quantified in terms of volume ($V_{\Theta_0}$); and with $\hat{\theta}_0$ randomly selected from this set.
	Smaller sets exhibit a more direct path between $x_0$ and the origin, which may be correlated with less prediction offset. 
	The influence of the size of the initial membership on the closed-loop cost is shown in Fig. \ref{fig:cost_big_thetas}, where the previous simulation was performed for forty random $\hat{\theta}_0$ for each set and then a box plot with the results of the IH closed-loop cost was drawn for each set.  For volumes from zero to approximately 0.33, it can be seen that for larger sets a worse performance may be expected, but also that a relatively good performance cannot be completely excluded for large sets. In the case of volumes larger than 0.33, the cost results get closer to each other. This may be due to the fact that the set of feasible inputs for the MPC optimization problem \eqref{eq:mpc_estimate} becomes significantly smaller as the membership set grows very large, because the system constraints must hold for all possible system trajectories.
	In the bound \eqref{eq:infinite-cost-bound}, $\Theta_0$ influences various scalars that characterize maximal changes of the system trajectory, for example $c_{A}$ included in $\Delta$, cf. \eqref{eq:prop1-final-ineq}. 
	The initial membership set also influences the bound of the MPC value function. However, a more precise quantitative assessment of the influence of this set is difficult to make.

	Fig. \ref{fig:traj_thetas} shows the closed-loop trajectory for different values of $\lVert\tilde{\theta}_0 \rVert$ and $\Theta_0 = \{\theta \in \R^3 \; | \; 0 \leq [\theta]_i \leq 0.75 \}$. Since there is an infinite number of $\tilde{\theta}_0$ for some specific norm, $\tilde{\theta}_0$ that point in the direction of $\theta^*$ were selected for the simulation. 
	Compared to the results for $\Theta_0$, the influence of $\tilde{\theta}_0$ is less pronounced in this simulation. In contrast to the case for $\Theta_0$, a precise statement about the influence of $\tilde{\theta}_0$ on the closed-loop performance can be made.
	The norm of the initial parameter error appears as a quadratic term in the performance bound of Thm. \ref{thm:performance_bound}, and thus a worst-case quadratic performance decrease can be expected. This is seen in the simulation results in Fig. \ref{fig:performance_vs_error}, where the worst-case closed-loop performance was plotted, which was determined as follows. For each value of the norm $\|\tilde{\theta}_0\|$, the system was simulated with different $\tilde{\theta}_0$ with this given norm and the highest closed-loop IH cost was selected.
	
	\begin{figure}
		\vspace{10pt}
		\centering
		\begin{tikzpicture}
			\begin{axis}[
				cycle list name=color list,xlabel={$[x_k]_1$}, ylabel={$[x_k]_2$}, grid=major, legend pos=north west,width=7.5cm,height=7cm,legend style={font=\small}],
				\addplot
				table[x=x,y=y,col sep=comma]{gfx/data2plot/different_thetas/trajectories1.csv};
				\addlegendentry{$0$};
				\addplot
				table[x=x,y=y,col sep=comma]{gfx/data2plot/different_thetas/trajectories2.csv};
				\addlegendentry{$0.25$};
				\addplot
				table[x=x,y=y,col sep=comma]{gfx/data2plot/different_thetas/trajectories3.csv};
				\addlegendentry{$0.5$};
				\addplot
				table[x=x,y=y,col sep=comma]{gfx/data2plot/different_thetas/trajectories4.csv};
				\addlegendentry{$0.75$};
				\addplot
				table[x=x,y=y,col sep=comma]{gfx/data2plot/different_thetas/trajectories5.csv};
				\addlegendentry{$1 $};
			\end{axis}
		\end{tikzpicture}
		\caption{Closed-loop trajectory for different $\|\tilde{\theta}_0\|$.}
		\label{fig:traj_thetas}
	\end{figure}
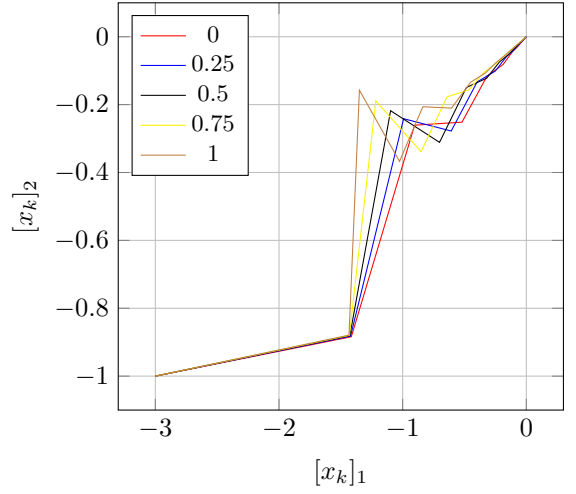		
	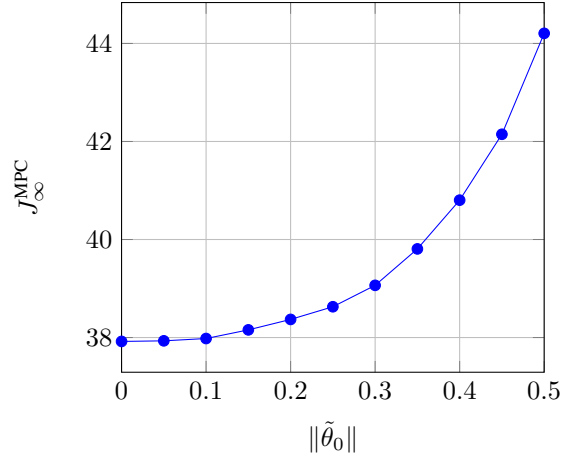
\begin{figure}
		\centering
		\begin{tikzpicture}
			\begin{axis}[
				xmin=0,xmax=0.5,xlabel={$\|\tilde{\theta}_0\|$}, ylabel={$J^{\text{MPC}}_\infty$}, grid=major, width=7.2cm,height=6.5cm,legend pos=north west],
				\addplot[mark=*,mark options={fill=blue},blue]
				table[x=x,y=y,col sep=comma]{gfx/data2plot/different_thetas/max_costs_worst_case.csv};
			\end{axis}
		\end{tikzpicture}
		\caption{Worst-case performance for different values of $\| \tilde{\theta}_0 \|$.}
		\label{fig:performance_vs_error}
	\end{figure}

	\section{Conclusion}
	\label{sec:conclusions}
	In this paper a bound for the closed-loop performance of an adaptive tube-based MPC scheme is proposed. 
	This bound was derived in two steps, first relating the value function of the MPC scheme to the optimal cost of the infinite horizon problem with true system parameters; and then using the difference of the MPC value function between two consecutive time steps to bound the infinite sum of the closed-loop cost. 
	The bound can be calculated a priori and is expressed in terms of various quantities associated with the control problem such as the initial system parameter error, the constraint set and parameters of the MPC scheme such as the weighting matrices among others. 
	The bound represents a first step into analyzing and quantifying the influence of online adaptation on the performance of MPC schemes, and thus into quantifying the benefits of online adaptation in comparison to other MPC schemes capable of handling uncertain models.

	\section*{APPENDIX}
	For $\bar{u}_{l|k}$, $\hat{x}_{l|k}(\bs{\bar{u}}_k;x_k)$, $k \in \N_0$, it holds $F\hat{x}_{l|k}(\bs{\bar{u}}_k;x_k)+G\bar{u}_{l|k}\leq 1,\; l\in \N_0^{N-1}$. 
	Then, $F\hat{x}_{l|k}(\bs{\bar{u}}_k;x_k)+G(u^*_{l|k}-\delta u_{l|k})\leq 1,\; l\in \N_0^{N-1}$, with $\delta u_{l|k}$ as previously defined. 
	An upper bound for $\lVert \delta u_{l|k} \rVert^2$ can be found using the solution of the optimization problem
	\begin{equation}
		\begin{split}
			\max_{\delta \bs{\bar{\nu}}_N,\bs{\nu}_N,\bs{\chi}_N} &\delta \bs{\bar{\nu}}_N^T \delta \bs{\bar{\nu}}_N \\
			\text{s.t.}\quad & F\hat{x}_{l|k}(\bs{\bar{u}}_k;x_k)+G(\nu_{l|k}-\delta \bar{\nu}_{l|k})\leq 1\\
			&F \chi_{l|k} + G \nu_{l|k} \leq 1, \quad \nu_{l|k}-\delta \bar{\nu}_{l|k} = \bar{u}_{l|k},\\
			&l \in \{0,\dotsc,N-1\}
		\end{split}
	\end{equation}
	where $\delta \bs{\bar{\nu}}_N:=[\delta \bar{\nu}_{1|k},\dotsc,\delta \bar{\nu}_{N-1|k}]$, and $\bs{\nu}_N$, $\bs{\chi}_N$ are defined in a similar manner. Notice that $\hat{x}_{l|k}(\bs{\bar{u}}_k;x_k)$ is the predicted trajectory for specific $x_k$, $\hat{\theta}_k$, $\Theta_k$ and $\bar{u}_{l|k}$ is the associated input. 
	Let $\delta \bs{\bar{u}}_N^*$ be the corresponding optimum, then it holds $\|\delta u_{l|k}\| \leq \|\delta \bar{u}_{l|k}^*\|$ for specific $x_k$, $\hat{\theta}_k$, $\Theta_k$.

	\bibliographystyle{plain}
	\bibliography{bib/refs}
	
\end{document}